\documentclass[pra,twocolumn,amsmath,amssymb]{revtex4}

\usepackage{bbm}
\usepackage{graphicx}
\usepackage{oldgerm}
\usepackage{amsmath}
\usepackage{amssymb}
\usepackage{amsbsy}
\usepackage{amsthm}
\usepackage{bm}
\usepackage[colorlinks=false]{hyperref}
\hypersetup{pdftitle={Experimentally Feasible Security Check for n-qubit Quantum Secret Sharing}}

\newcommand{\beq}{\begin{eqnarray}}
\newcommand{\eeq}{\end{eqnarray}}

\newcommand{\bra}[1]{\langle #1|}
\newcommand{\ket}[1]{|#1 \rangle}

\newcommand{\oP}[2]{\ket{#1}\bra{#2}}
\newcommand{\I}{\mathbb{I}}

\begin{document}

\title{Experimentally Feasible Security Check for $n$-qubit Quantum Secret Sharing}

\author{Stefan Schauer}
\affiliation{AIT Austrian Institute of Technology GmbH, Donau-City-Str. 1, A-1220 Vienna, Austria}
\author{Marcus Huber}
\affiliation{Faculty of Physics, University of Vienna, Boltzmanngasse 5, A-1090 Vienna, Austria}
\author{Beatrix C. Hiesmayr}
\affiliation{Research Center for Quantum Information, Institute of
Physics, Slovak Academy of Sciences, Dubravska cesta 9, 84511
Bratislava, Slovakia}
\affiliation{Faculty of Physics, University of Vienna, Boltzmanngasse 5, A-1090 Vienna, Austria}

\begin{abstract}
In this article we present a general security strategy for quantum secret sharing (QSS) protocols based on the HBB scheme presented by Hillery, Bu\v{z}ek and Berthiaume [Phys. Rev A \textbf{59}, 1829 (1999)]. We focus on a generalization of the HBB protocol to $n$ communication parties thus including $n$-partite GHZ states. We show that the multipartite version of the HBB scheme is insecure in certain settings and impractical when going to large $n$. To provide security for such QSS schemes in general we use the framework presented by some of the authors [M. Huber, F. Minert, A. Gabriel, B. C. Hiesmayr, Phys. Rev. Lett. \textbf{104}, 210501 (2010)] to detect certain genuine $n$ partite entanglement between the communication parties. In particular, we present a simple inequality which tests the security.
\end{abstract}

\maketitle

\section{Introduction}

In classical cryptography secret sharing has been introduced by Shamir \cite{Sha79} and Blakley \cite{Bla79} in 1979 and is useful in many applications. The main idea is to divide a secret into several shares and distribute these shares among several parties such that the secret can be reconstructed when a certain number of parties (or all) come together and combine their shares. Additionally, each party alone is not able to gain any information about the secret. The idea of secret sharing has been brought to Quantum Cryptography in 1999 when Hillery, Bu\v{z}ek and Berthiaume presented their scheme \cite{HBB99} based on GHZ states. Since then Quantum Secret Sharing (QSS) has been another field of great interest besides Quantum Key Distribution (QKD). In the same year Karlsson, Koashi and Imoto also presented a similar QSS protocol based on Bell states \cite{KKI99} and several other schemes followed \cite{CGL99,Got00,TZG01,GG03b,LZP04,DLLZ05a,DLLZ05b,DLLZ06,GKBW07,MS08b,CCKKL08,HWB08,KMMP09}.
\par
Most of these protocols make heavy use of entangled states to communicate between several parties. In general, the security of such protocols is rather complex to analyze since there are more parties involved compared to QKD and some of the legal participants have to be considered dishonest. This model of adversaries from the inside is in fact much stronger because such an adversary in general has more advantages than an eavesdropper from the outside. The success of the protocol depends strongly on the fact that all parties share a certain genuine multipartite entangled state after transmission. We show in this paper that the security of a protocol can be obtained by checking for this certain genuine multipartite entanglement. For that we use the framework presented in Refs.~\cite{HMGH10,HSSG10} which provide Bell-like inequalities which are experimentally testable.
\par
In the following section we shortly review the HBB scheme including the argument presented in Ref.~\cite{QGWZ07} regarding the security against a cheating Charlie. Further, we discuss the generalization of the HBB scheme to $n$ qubits and present a successful eavesdropping strategy based on the argument in Ref.~\cite{QGWZ07}. Based on the inequalities we provide a new security argument for $n$ qubit secret sharing protocols.

\section{The HBB Scheme} \label{sec:HBBScheme}

In their article \cite{HBB99} Hillery, Bu\v{z}ek and Berthiaume presented a quantum secret sharing scheme based on the distribution of GHZ states of the form
\begin{equation}
\ket{\Psi_0} = \frac{1}{\sqrt{2}} (\ket{000} + \ket{111})_{ABC}
\label{eq:GHZState}
\end{equation}
between three parties, Alice, Bob and Charlie. Each party measures its qubit at random in one of two bases. Based on their results, Bob and Charlie together are able to determine Alice's result but individually have no information about it.
\par
In detail, Alice generates copies of the state $\ket{\Psi}$ in her laboratory and sends qubit $B$ to Bob and qubit $C$ to Charlie. Then, each party randomly chooses to measure its qubit either in the $X$ or in the $Y$ basis. The eigenstates of these bases are
\begin{equation}
\ket{x^\pm} = \frac{1}{\sqrt{2}} (\ket{0} \pm \ket{1})
\quad
\ket{y^\pm} = \frac{1}{\sqrt{2}} (\ket{0} \pm i\ket{1}).
\end{equation}
Taking the $X$ basis the GHZ state $\ket{\Psi}$ can be written as
\begin{equation}
\begin{aligned}
\ket{\Psi}_{ABC} = \frac{1}{2} [(&\ket{x^+}_A\ket{x^+}_B + \ket{x^-}_A\ket{x^-}_B) \ket{x^+}_C \\
+ (&\ket{x^+}_A\ket{x^-}_B + \ket{x^-}_A\ket{x^+}_B) \ket{x^-}_C].
\end{aligned}
\label{eq:StateBasisX}
\end{equation}
From this fact it is easy to see that if both Alice and Bob perform their measurements in the $X$ basis and obtain the same result, Charlie ends up with the state $\ket{x^+}$. Otherwise, if Alice and Bob obtain different results, Charlie ends up with the state $\ket{x^-}$. Regarding the case when Alice and Bob perform their measurement both in the $Y$ basis or in different bases similar conditions can be found for Charlie's state (c.f. table \ref{tab:MeasResults}).

\begin{table}{\vspace{0.5cm}}
\begin{center}
\begin{tabular}{cc}
 & Alice \\ Bob &
\setlength{\tabcolsep}{3mm}
\begin{tabular}{c|cccc}
            & $\ket{x^+}$ & $\ket{x^-}$ & $\ket{y^+}$ & $\ket{y^-}$ \\ \hline
$\ket{x^+}$ & $\ket{x^+}$ & $\ket{x^-}$ & $\ket{y^+}$ & $\ket{y^-}$ \\
$\ket{x^-}$ & $\ket{x^-}$ & $\ket{x^+}$ & $\ket{y^-}$ & $\ket{y^+}$ \\
$\ket{y^+}$ & $\ket{y^-}$ & $\ket{y^+}$ & $\ket{x^-}$ & $\ket{x^+}$\\
$\ket{y^-}$ & $\ket{y^+}$ & $\ket{y^-}$ & $\ket{x^+}$ & $\ket{x^-}$
\end{tabular}
\end{tabular}
\end{center}
\caption{Charlie's state depending on Alice's and Bob's measurement result.}
\label{tab:MeasResults}
\end{table}

After each party performed its measurement they all announce their bases for the whole sequence sent by Alice but do not reveal the specific result. Additionally, all three parties sacrifice some of the remaining measurement results to check for eavesdroppers and dishonest parties by comparing them publicly. Based on the information about the basis choice of the remaining qubits Charlie always knows whether Alice and Bob have the same results or not, but he has no information about their exact results. Further, Bob knows that he either has the same or the opposite result of Alice and thus needs the information about Charlie's measurement result to fully determine it. Thus, Bob and Charlie have to collaborate to obtain Alice's result. Due to the random choice of the measurement bases, Charlie will measure in the wrong basis half of the times. These cases can be identified when the three parties reveal their bases and the respective qubits have to be discarded.
\par
The security argument, as is described above, has been presented in Ref.~\cite{HBB99} but later that year Karlsson et al. commented on the HBB scheme that the order in which the measurement bases and the results for the test bits are revealed is crucial \cite{KKI99}. They showed that the HBB scheme becomes insecure if the measurement bases are revealed before the results for the test bits. They suggested the following sequence: first, Bob and Charlie publicly disclose their measurement results for the test bits and afterwards, in the reversed order, they announce the corresponding measurement bases. The reversed order is important such that none of them can gain too much information from the actions of the previous parties. We want to stress that this is, nevertheless, not a very efficient way to secure the protocol since the order of the messages is not implicitly preserved by the network. Alice has to tell each party when to send its result and has to wait on the response. In case of three parties as in the HBB scheme there is no big difference but it can become a large overhead when going to $n$ parties.

\section{A new Security Argument}

In three articles \cite{HMGH10,GHH10,HSSG10} the authors presented a series of inequalities to test for genuine multipartite entanglement and for k-separability for any multipartite qudit system. These Bell-like inequalities are easily experimentally implementable as only local observables are needed. We present here how two inequalities designed for the HBB protocol described above can be used to check for adversaries.
\par
The idea is that the attack strategy based on auxiliary qubits as presented in Ref.~\cite{KKI99} does not work if the parties can verify that they share a genuinly multipartite entangled $n$-qubit state. The intervention of an untrusted party, e.g. Charlie, is based on the auxiliary qubits he introduces into the protocol to gain additional information about Bob's results. Differently stated, it changes the overall state and this can be detected by performing certain additional setups and evaluating the inequalities given in eq. (\ref{eq:Ineq3Qubit}) below.
\par
Before we present the inequalities we need to define bi-separability: If the density operator of a $3$ qubit state can be decomposed into the following form
\begin{eqnarray} &&\rho_{bisep}=\nonumber\\
&&\sum_j p_j \rho_{AB}^j\otimes\rho_C^j+\sum_j q_j
\rho_{AC}^j\otimes\rho_B^j+\sum_j r_j
\rho_{BC}^j\otimes\rho_A^j\;,\nonumber\\
\end{eqnarray} with $ p_j,q_j,r_j\geq
0$ and $\sum_j p_j+q_j+r_j=1$, it is called biseparable. Here the two-body states $\rho^j_{AB}$, $\rho^j_{BC}$ and $\rho^j_{AC}$ can describe entangled states. Even though there is no bipartite splitting with respect to which the state $\rho$ is separable, it is considered biseparable since it can be prepared through a statistical mixture of bipartite entangled states. Generalization to $n$-qubit states is straightforward.
\par
Based on the bi-separability we can define the inequalities for the 3-qubit case of the HBB protocol. Using $\sigma_1:=\mathbb{I}$ and the
abbreviation for the Pauli operators
\begin{equation}
\langle\sigma_i\otimes\sigma_j\otimes\sigma_k\rangle_\rho:=ijk
\end{equation}
we can rewrite and linearize the inequalities derived in Refs.~\cite{HMGH10,HSSG10} in terms of local observables:
\begin{widetext}
\begin{eqnarray}
I_1:\quad\frac{1}{8} (xxx - yyx - yxy - xyy) - \frac{1}{16}(3\cdot111 - zz1 - z1z - 1zz) &\leq& 0\; \nonumber \\
I_2:\quad\frac{1}{8} (yyy - xxy - xyx - yxx) -
\frac{1}{16}(3\cdot111 - zz1 - z1z - 1zz) &\leq& 0\;.
\label{eq:Ineq3Qubit}
\end{eqnarray}
\end{widetext}
These inequalities are satisfied for all biseparable states. It is convex, therefore it obviously valid for mixed states.
\par
As it is easy to see the first inequality uses combinations of local observables which are needed in the original HBB scheme to form the secret key (cf. table \ref{tab:MeasResults}) whereas the second inequality uses combinations which are discarded in the original protocol (i.e. $yyy$, $yxx$, $xyx$ and $xxy$). Unfortunately, the latter one can only be applied if the initial state is the ''imaginary'' GHZ state
\begin{equation}
\ket{\Phi_0} = \frac{1}{\sqrt{2}} (\ket{000} + i\ket{111}).
\end{equation}
Thus, we have to adjust the original HBB protocol in the following way: Alice prepares at random one of two states, either the standard GHZ state $\ket{\Psi}$ or the state $\ket{\Phi}$. Then, she distributes the qubits between Bob and Charlie as in the original protocol. Due to the use of the inequalities in eq. (\ref{eq:Ineq3Qubit}) the $Z$-basis has to be introduced as an additional measurement basis. After Bob and Charlie performed their measurements they announce their bases and Alice tells them to reveal some of their results to test for the inequalities. Here, Alice tests with the first inequality of eq. (\ref{eq:Ineq3Qubit}) whenever she prepared the state $\ket{\Psi}$ and with the second inequality when she prepared $\ket{\Phi}$. We want to stress that Alice does not announce which initial state she prepared until after the check for eavesdroppers. Therefore, the sequence in which Bob and Charlie announce their bases and results is irrelevant since a cheating Charlie can not be sure whether Alice initially prepared $\ket{\Psi}$ or $\ket{\Phi}$. Hence, Charlie introduces a certain error and will be detected by the legitimate parties as it is explained in detail in the next section.
\par
The application of the inequalities makes it possible to overcome the check for the correct order of the messages and thus makes the protocol less complex. The introduction of the second GHZ state $\ket{\Phi}$ does not influence the efficiency, since combinations of observables which are discarded in the original protocol can be used with the state $\ket{\Phi}$ and vice versa. The only drawback is the additional measurement basis $Z$, which is not necessary to establish the secret but is needed to compute the inequalities. Fortunately, we can overcome also this problem by choosing $Z$ only with a certain probability $q$, which can go to 0 in the asymptotic limit.

\section{Security proof for $3$ qubits}

In particular the first inequality in eq. (\ref{eq:Ineq3Qubit}) is violated by the GHZ-state $\ket{\Psi}$ in the computational basis with the value $\frac{1}{2}$, which is the optimum for any GHZ-state representation. Note that there are several representations of the GHZ-state which would give no violation.
\par
The security check -- optimized for the basis system the three parties agreed on -- would therefore use some of the measurement results to evaluate the inequalities (similar to the check for adversaries suggested in \cite{HBB99}). Additionally, the three parties also have to perform measurements in the $Z$ basis to evaluate the inequalities, which slightly changes the protocol, as pointed out above. If the inequalities are violated, the parties can be sure that no adversary is present, what we prove in the following.
\par
In Ref.~\cite{QGWZ07} it has been shown -- using a more general
approach than in \cite{KKI99} -- that the original HBB scheme
\cite{HBB99} is insecure against a dishonest Charlie. The main idea
is again that Charlie intercepts the qubit flying to Bob and
entangles it with an ancillary qubit. Later on, he uses his qubit
together with the ancillary qubit to infer Alice's measurement
result without Bob's assistance. In detail, Charlie uses an
ancillary qubit in the state $\ket{0}_E$ and entangles it with the
intercepted qubit $B$ using the Hadamard operation
$H=1/\sqrt{2}(\oP{0}{0}+\oP{0}{1}+\oP{1}{0}-\oP{1}{1})$ on qubit $B$
and a CNOT operation $\text{CNOT} = \oP{0}{0} \otimes \I + \oP{1}{1}
\otimes \sigma_x$ on qubits $B$ and $E$. This brings the initial
system $\ket{\Psi_0}_{ABC} \otimes \ket{0}_{E}$ into the states
\begin{equation}
\ket{\Psi_1} = \frac{1}{2} (\ket{0000} + \ket{0101} + \ket{1010} -
\ket{1111})_{ABCE}\;. \label{stateABCE}
\end{equation}
Charlie sends qubit $B$ to Bob and waits until Alice and Bob announce their measurement bases. According to the measurement results of Alice and Bob, the qubits $C$ and $E$ in Charlie's possession collapse into some predefined state. In case both Alice and Bob measure in the $X$ basis Charlie obtains one of the states
\begin{equation}
\begin{aligned}
\ket{\Psi_{x^+x^+}} &= \frac{1}{2} (\ket{00} + \ket{01} + \ket{10} - \ket{11})_{CE} \\
\ket{\Psi_{x^+x^-}} &= \frac{1}{2} (\ket{00} - \ket{01} + \ket{10} + \ket{11})_{CE} \\
\ket{\Psi_{x^-x^+}} &= \frac{1}{2} (\ket{00} + \ket{01} - \ket{10} + \ket{11})_{CE} \\
\ket{\Psi_{x^-x^-}} &= \frac{1}{2} (\ket{00} - \ket{01} - \ket{10} -
\ket{11})_{CE}\;.
\end{aligned}
\label{eq:CharliesState}
\end{equation}
Charlie uses this fact together with the information about Bob's
measurement basis and result to determine the correct value he has
to announce to stay undetected. Further, Charlie is also able to
compute Alice's result without any help of Bob \cite{QGWZ07}, which
makes the whole protocol insecure.
\par
Taking our suggested modified version of the HBB scheme performing
the check for adversaries based on the inequalities and employing 2
GHZ states at random Charlie is always detected with a certain
probability. As pointed out, Charlie's attack mainly relies on the
information about Bob's bases and results, which he can also obtain
in our modified version. Nevertheless, Charlie is unable to decide
which initial state Alice prepared such that he can only guess the
correct result to violate the inequalities.
\par
In detail after Charlie's attack the four qubit state is a mixture
of the two following states are mixed
\begin{eqnarray}
\ket{\Psi_1} &=& \frac{1}{2} (\ket{0000} + \ket{0101} + \ket{1010} -
\ket{1111})_{ABCE}\nonumber\\ \ket{\Phi_2} &=& \frac{1}{2}
(\ket{0000} + \ket{0101} + i \ket{1010} -i
\ket{1111})_{ABCE}\;.\nonumber\\ \label{stateABCE2}
\end{eqnarray}
Ignoring Charlies additional qubit the first inequality derives to
$I_1: -\frac{1}{2}-p\leq 0$ and the second inequality derives to
$I_2: \frac{1}{2}-p\leq 0$ with $p$ being the probability that Alice chooses the state $\ket{\Psi_1}$. 
These values are different to the expected values without
cheating parties, thus Charlie will be revealed. On the other hand
Charlie can try to act by local unitaries on qubit $C$ or by
unitaries on qubits $CE$ (here we used the convenient
parametrization of the unitary group $U(4)$ in Ref.~\cite{SHH2})
such that the value of $I_1$ gets more positive but the trade off is
that $I_2$ gets more negative, again this can be detected. In
summary, the suggested attack to the HBB scheme presented in
Ref.~\cite{QGWZ07} as well as any generalization of it is detected
by the test of the two inequalities.

\section{Security Proof for $n$ qubits}

The inequalities provided by the framework presented in Refs.~\cite{HMGH10,HSSG10} can be extended to any number of qubits. To give an example, for the 4-qubit case described in the previous section we get a similar inequalities:
\begin{widetext}
\begin{eqnarray}
\frac{1}{8}(xxxx - yyxx - yxyx - xyyx - xxyy - xyxy - yxxy + yyyy) &-& \nonumber \\
\frac{1}{16}(7\cdot1111 - zz11 - z11z - 11zz - z1z1 - 1z1z - 1zz1 - zzzz) &\leq& 0 \qquad \text{and} \nonumber \\
\frac{1}{8}(xxxy + xxyx + xyxx + yxxx - xyyy - yxyy - yyxy - yyyx) &-& \nonumber \\
\frac{1}{16}(7\cdot1111 - zz11 - z11z - 11zz - z1z1 - 1z1z - 1zz1 - zzzz) &\leq& 0
\label{eq:Ineq4Qubit}
\end{eqnarray}
\end{widetext}
Also in this case, the four communicating parties sacrifice some of their measurement results to test the inequalities. If they are satisfied they have to assume that an adversary is present. Extending the attack strategy from \cite{GKBW07} to four parties the state Charlie uses in his attack is a 6-qubit entangled state. Due to his intervention, the genuine 4-qubit state is destroyed and thus no genuine 4-qubit entanglement can be detected using the inequalities. Hence, the legitimate communication parties discover Charlie's intervention and abort the protocol.
\par
The inequalities for $n$ qubits can be derived straight forward from the 3-qubit case (eq. \ref{eq:Ineq3Qubit}) and the 4-qubit case (eq. \ref{eq:Ineq4Qubit}). Thus, the check for adversaries can be performed in the same way as described above. This gives the advantage that the communication parties no longer have to rely on the order of the messages. The adversary, Charlie, does not know which of the measurement results count for the test of the inequalities and which count for the secret. Hence, he sends measurement results which do not violate the inequalities and therefore he is detected by the other parties also in the most general case.

\section{Conclusion} \label{sec:Conclusion}

In this article we presented a security argument for general
HBB-type quantum secret sharing schemes between $n$ parties. The
check for adversaries of such protocols is in general getting more
and more inefficient if a large number of parties is involved. We
present a different security strategy based on the verification of
genuine multipartite entanglement itself which is at the heart of
such protocols and in addition that is efficient also for large
number of parties.
\par 
In a slightly different version of the HBB protocol we described a
way to integrate this security check efficiently, i.e. by simple
Bell--like inequalities (Refs.~\cite{HMGH10,HSSG10}) adapted to the
protocol. They use the data which is usually not regarded for the
protocol and a measurement in a third direction has to be
introduced.
\par 
A test of these inequalities is a much stronger statement than the
common test for eavesdroppers presented e.g. in
Refs~\cite{HBB99,KKI99} as they indicate the presence of an
adversary because any adversary has to change the $n$-partite
entangled state in order to obtain any information on the secrete.
\par 
Certainly, our presented general scheme may be applied to secret
sharing protocols involving multi-qudit systems or graph states.

\section{Acknowledgements}
We want to thank Martin Suda, Christoph Spengler, Andreas Gabriel and Jan Bouda for interesting discussions and fruitful comments on our work. S. Schauer would like to thank the AIT Austrian Institute of Technology GmbH for it's support.  M. Huber gratefully acknowledges the support of the Austrian Fund project FWF-P21947N16.

\bibliography{nQSS}

\end{document}